\documentclass[%
 reprint,
superscriptaddress,
 amsmath,amssymb,
 aps,
 longbibliography,
]{revtex4-2}

\usepackage{dcolumn}
\usepackage{bm}
\usepackage{comment}

\usepackage{amsmath,amsfonts}
\usepackage{bm}
\usepackage{mathcomp}
\usepackage{physics}
\usepackage[dvipdfmx]{color}
\usepackage{float}
\usepackage{tikz}
\usetikzlibrary{intersections, calc, arrows.meta}
\usepackage{ascmac}
\usepackage{fancybox}

\usepackage{ulem}

\usepackage{hyperref}

\begin{document}

\preprint{APS/123-QED}

\title{Topological phase diagram of the Haldane model \\ on a Bishamon-kikko--honeycomb lattice}

\author{Sogen Ikegami}
  \email{ikegami-sogen443@g.ecc.u-tokyo.ac.jp}
  \affiliation{Department of Applied Physics, The University of Tokyo, Tokyo 113-8656, Japan}
\author{Kiyu Fukui}
  \affiliation{Department of Applied Physics, The University of Tokyo, Tokyo 113-8656, Japan}
\author{Shun Okumura}
  \affiliation{Department of Applied Physics, The University of Tokyo, Tokyo 113-8656, Japan}
\author{Yasuyuki Kato}
  \affiliation{Department of Applied Physics, University of Fukui, Fukui 910-8507, Japan}
\author{Yukitoshi Motome}%
 \email{motome@ap.t.u-tokyo.ac.jp}
  \affiliation{Department of Applied Physics, The University of Tokyo, Tokyo 113-8656, Japan}

\date{\today}

\begin{abstract}
Topological flat bands have gained extensive interest 
as a platform for exploring the interplay between nontrivial band topology and correlation effects. 
In recent studies, strongly correlated phenomena originating from a topological flat band were discussed on
a periodically 1/6-depleted honeycomb lattice, 
but the fundamental topological nature associated with this lattice structure remains unexplored.
Here we study the band structure and topological phase diagram for the Haldane model on this lattice, 
which we call the Bishamon-kikko lattice. 
We also extend our study to the model connecting the Bishamon-kikko and honeycomb lattices.
We show that these models exhibit richer topological characteristics 
compared to the original Haldane model on the honeycomb lattice, such as 
topological insulating states with higher Chern numbers,
metallic states with nontrivial band topology even at commensurate electron fillings, and metal-insulator transitions between them.
Our findings offer a playground of correlated topological phenomena and stimulate their realization in a variety of two-dimensional systems, 
such as van der Waals materials, graphene nanostructures, and photonic crystals. 
\end{abstract}

\maketitle


\section{\label{sec1}Introduction}

Topological phase of matter is one of the most exciting discoveries in modern condensed matter physics. 
This concept was originally brought by the quantum Hall effect~\cite{PhysRevLett.45.494,PhysRevLett.50.1395,PhysRevLett.49.405}, 
and later has opened a variety of exotic states, such as topological insulators~\cite{PhysRevLett.95.226801,doi:10.1126/science.1148047,PhysRevB.75.121306} 
and topological superconductors~\cite{PhysRevB.61.10267,A_Yu_Kitaev_2001,PhysRevLett.100.096407}. 
In these systems, the topological invariant called the Chern number $C$~\cite{PhysRevLett.49.405,KOHMOTO1985343}, 
which is given by summation of the Berry curvature~\cite{Berry1984} across the Brillouin zone for each band, 
plays a pivotal role in the characterization of the topological nature. 
For instance, the Haldane model on a honeycomb lattice exhibits topological phases with $C=\pm 1$, 
leading to a spontaneous quantized anomalous Hall effect~\cite{PhysRevLett.61.2015}. 
Besides, flat bands, which are dispersionless bands with a constant energy in momentum space, 
have also attracted much interests. 
A typical example can be found in a tight-binding model on the so-called Lieb lattice, 
a square lattice with $1/4$ defects, where electrons are localized due to quantum interferences around the defects~\cite{PhysRevLett.62.1201}. 
In such flat band systems, effects of electron correlations are extremely enhanced 
and lead to instabilities toward various strongly correlated phenomena, 
such as ferromagnetism~\cite{A_Mielke_1991_1,A_Mielke_1991_2,A_Mielke_1992,PhysRevLett.69.1608,PhysRevLett.109.096404} and
high-temperature superconductivity~\cite{PismaZhETF.51.488,Tang2014}.
Moreover, topological flat bands, i.e., nearly flat bands with nonzero Chern numbers, 
have gained significant attention since they provide a platform for realizing the fractional quantum Hall states, 
where non-Abelian fractional excitations can be utilized for topological quantum computing~\cite{PhysRevLett.106.236802,PhysRevLett.106.236803,PhysRevLett.106.236804,doi:10.1142/S021797921330017X}.

Recently, a topological flat band was discussed for a honeycomb version of the Lieb lattice 
obtained by periodically depleting 1/6 sites from the honeycomb lattice~\cite{doi:10.1126/sciadv.adg0028,chen2022emergent}. 
This lattice structure belongs to a class of superhoneycomb systems~\cite{PhysRevLett.71.4389}, 
and was recently found in quenched van der Waals ferromagnet Fe$_{5-\delta}$GeTe$_2$~\cite{Wu2024,A.F.May2019}.
The previous studies showed that electron correlations on this lattice can bring about 
orbital-selective Mott transitions and a topological Kondo semimetal~\cite{chen2022emergent,doi:10.1126/sciadv.adg0028}. 
Thus, this lattice has potential for exotic phenomena associated with topological flat bands,
but the fundamental topological nature has not yet been fully explored.

In this paper, we investigate the band structure and topological nature 
of the prototypical Haldane model on the $1/6$-depleted honeycomb lattice. 
While this lattice structure was called a clover lattice~\cite{chen2022emergent}, 
we here call it the Bishamon-kikko (BK) lattice after Japanese traditional patterns, 
since the clover lattice often refers a group of structures with $C_4$ or $C_2$ rotational symmetry; 
the BK lattice has $C_3$ rotational symmetry as shown below. 
Calculating the Berry curvature for the occupied bands, we systematically  
map out the topological phase diagrams while changing the model parameters and electron filling. 
In addition, we consider a connection between the BK lattice and the original honeycomb lattice 
by continuously changing the transfer integrals around the defect sites, 
which we call the Bishamon-kikko--honeycomb (BKH) lattice.
We find that these models exhibit richer topological properties not seen in the original Haldane model on the honeycomb lattice, 
such as topological insulating states with higher Chern numbers, metallic states 
with nontrivial band topology even at commensurate filling, and metal-insulator transitions between them. 

The organization of this paper is as follows.
In Sec.~\ref{sec2}, we introduce our models on the BK lattice and the BKH lattice, 
and the method to clarify the topological nature of the systems.
In Sec.~\ref{sec3}, we present the results for band structures and topological phase diagrams for both models.
Finally, Sec.~\ref{sec4} is devoted to a summary.

\section{\label{sec2}Model and method}

\begin{figure}[b]
  \includegraphics[width=\columnwidth]{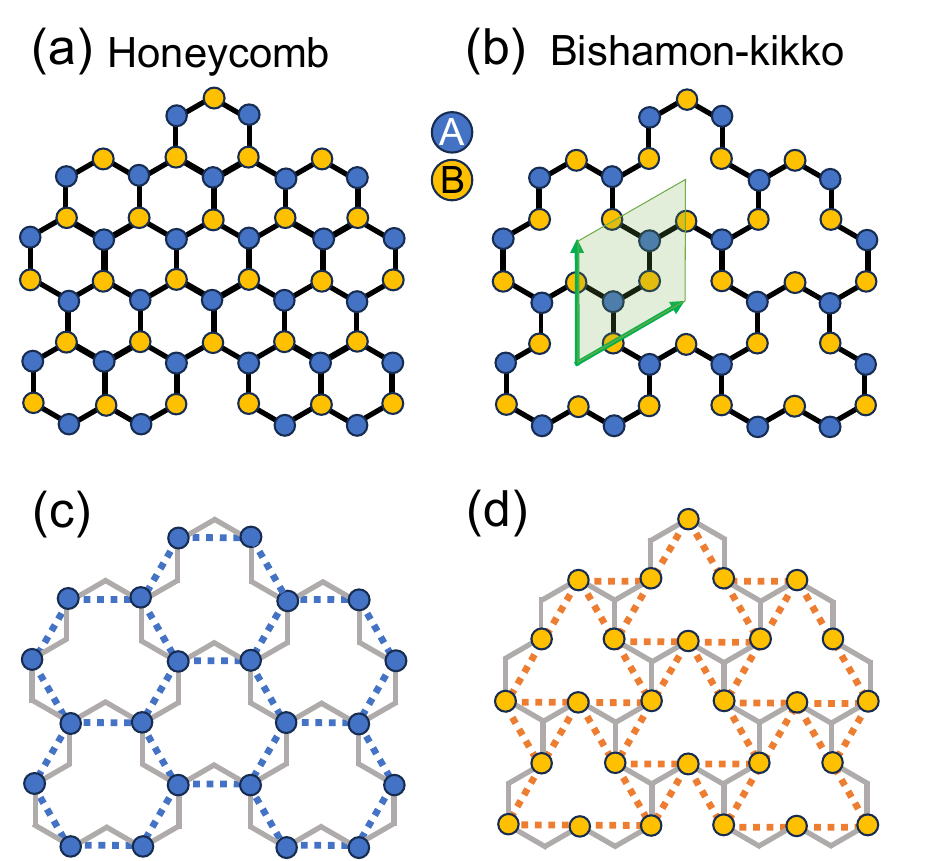}
  \caption{\label{fig:schematic} Schematic pictures of (a) the honeycomb and (b) BK lattices considered in the present study. 
  The blue and yellow circles denote two sublattices, A and B, respectively.
  In (b), the green arrows indicate the primitive translation vectors, and the green rhombus shows the unit cell.
  Schematics of the limits of (c) $M\rightarrow \infty$ and (d) $M\rightarrow -\infty$ of the Haldane model on the BK lattice [Eq.~\eqref{eq:h1}].
  The dashed lines represent the second-neighbor hopping surviving in each limit, and the gray lines are guides for the eye.
  The limit in (c) results in a honeycomb lattice of the second-neighbor hopping, while that in (d) is equivalent to a kagome lattice.
}
\end{figure}

\subsection{\label{subsec2-1}Bishamon-kikko lattice}
First, we introduce the Haldane model~\cite{PhysRevLett.61.2015} on the BK lattice 
obtained by periodically depleting $1/6$ lattice sites of the honeycomb lattice, as shown in Figs.~\ref{fig:schematic}(a) and \ref{fig:schematic}(b). 
While there are several types of the BK lattice depending on the way of site depletion 
as represented by different Japanese traditional pattens, 
we here focus on the type shown in Fig.~\ref{fig:schematic}(b). 
Note that the BK lattice is similar but different from a decorated honeycomb lattice, which is reported to have a flat band and higher-order topological insulator phase \cite{PhysRevA.102.033527,doi:10.1021/acs.nanolett.0c03049,doi:10.1021/acsphotonics.1c01171}.
We consider the Haldane model on this BK lattice, whose Hamiltonian is given by
\begin{equation}\label{eq:h1}
  \hat{H} = \sum_{i}M_{i}\hat{c}_{i}^{\dagger}\hat{c}_{i}+ \left[\sum_{\langle i,j \rangle }t_{1}\hat{c}^{\dagger}_{i}\hat{c}_{j} + \sum_{\langle \langle i,j  \rangle\rangle }t_{2}e^{i\phi_{ij}}\hat{c}^{\dagger}_{i}\hat{c}_{j} + {\rm h.c.}\right],
\end{equation}
where $\hat{c}_{i}$ and $\hat{c}_{i}^{\dagger}$ are the annihilation and creation operators of spinless fermions at site, respectively. 
The summations $\langle i,j  \rangle$ and $\langle \langle i,j \rangle \rangle$ are taken for all nearest-neighbor pairs 
and second-neighbor pairs connected by two nearest-neighbor bonds, respectively.
Here, $M_{i}$ denotes a staggered site potential, taking $-M$ and $+M$ on the A and B sublattices respectively 
represented by the blue and yellow circles in Fig.~\ref{fig:schematic}(b);
$t_{1}$ and $t_{2}$ represent the transfer integrals for the nearest-neighbor and second-neighbor hopping, 
and $\phi_{ij}$ denotes the phase for the latter, which takes $+\phi$ and $-\phi$ 
for the hopping in the clockwise and counterclockwise directions with respect to the center of each plaquette, respectively. 
We note that, in the limit of $M\rightarrow \infty \ (-\infty)$, the yellow (blue) sites are projected out, 
and the model ends up with the honeycomb (kagome) lattice model 
composed of the second-neighbor hopping in the original model when the electron filling is less than 2/5 (3/5), 
as illustrated in Fig.~\ref{fig:schematic}(c) [\ref{fig:schematic}(d)]. 

\subsection{\label{subsec2-2}Bishamon-kikko--honeycomb lattice}
Next, we introduce the Haldane model on a lattice connecting the BK and honeycomb lattices. 
For this purpose, we introduce a parameter $\alpha$ that controls the transfer integrals  
including defects sites on the BK lattice,
and define the Hamiltonian on the honeycomb lattice as 
\begin{equation}\label{eq:h2}
  \begin{split}
  \hat{H} = &\sum_{i}M_{i}\hat{c}_{i}^{\dagger}\hat{c}_{i} \\ 
  &+\left[\sum_{\langle i,j \rangle }\alpha_{ij}t_{1}\hat{c}^{\dagger}_{i}\hat{c}_{j} + \sum_{\langle \langle i,j  \rangle\rangle }\alpha_{ij}t_{2}e^{i\phi_{ij}}\hat{c}^{\dagger}_{i}\hat{c}_{j} + {\rm h.c.}\right], 
  \end{split}
\end{equation}
where $\alpha_{ij} = \alpha$ ($0 \leq \alpha \leq 1$) when the hopping process includes a defect site on the BK lattice 
(for the nearest-neighbor case, when $i$ or $j$ belongs to a defect site; for the second-neighbor case, 
when two neighboring bonds consisting the second-neighbor hopping includes a defect site), 
and otherwise $\alpha_{ij}=1$; 
the other parameters are the same as in Eq.~(\ref{eq:h1}). 
In Eq.~\eqref{eq:h2}, the summations are taken on the honeycomb lattice. 

The system is equivalent to the Haldane model on the honeycomb lattice when $\alpha=1$. 
On the contrary, $\alpha=0$ dispatches the defect sites, indicating that the model is reduced to 
Eq.~\eqref{eq:h1} on the BK lattice with isolated defects.
Thus, Eq.~\eqref{eq:h2} smoothly connects the two lattice systems, and 
we call the lattice structure the ``Bishamon-kikko--honeycomb (BKH) lattice" in the following.

\subsection{\label{subsec2-3}Method}
We investigate the band structures and the topological nature of the Haldane models on the BK lattice 
and the BKH lattice. 
To characterize the topological nature,
we calculate the summation of the Berry curvature for the occupied states as 
\begin{equation}\label{eq:Chern}
  C_{\rm occ} = \frac{1}{2\pi} \sum_{n,\bf{k}}^{{\rm occ}} \left[\nabla_{\bm{k}} \times \bm{A}_{n\bm{k}}\right]_{k_{z}},
\end{equation}
where $\bm{A}_{n\bm{k}}=-i\expectationvalue{\nabla_{\bm{k}}}{u_{n\bm{k}}}$ is the Berry connection; 
$|u_{n\bm{k}}\rangle$ denotes the eigenstate at band $n$ with momentum $\bm{k}$,
and the subscript $k_{z}$ represents the $z$ component of $\nabla_{\bm{k}} \times \bm{A}_{n\bm{k}}$.
The summation in Eq.~\eqref{eq:Chern} is taken for all the occupied states below the Fermi level. 
Thus, $C_{\rm occ}$ becomes an integer called the Chern number $C$ when the system is insulating; 
a nonzero quantized value of $C_{\rm occ}$ indicates that the system is a topologically nontrivial insulator.

In the actual calculations of $C_{\rm occ}$, we adopt the Fukui-Hatsugai-Suzuki method 
that facilitates precise numerical estimates of the Berry curvature~\cite{doi:10.1143/JPSJ.74.1674}. 
Dividing the hexagonal Brillouin zone into $7500$ rhombi, 
we calculate the Berry curvature for each plaquette and add them up to calculate $C_{\rm occ}$.
Hereafter, we set $t_{1}=1$ as the energy scale. 

\section{\label{sec3}Results}

\begin{figure}[thp]
 \includegraphics[width=\columnwidth]{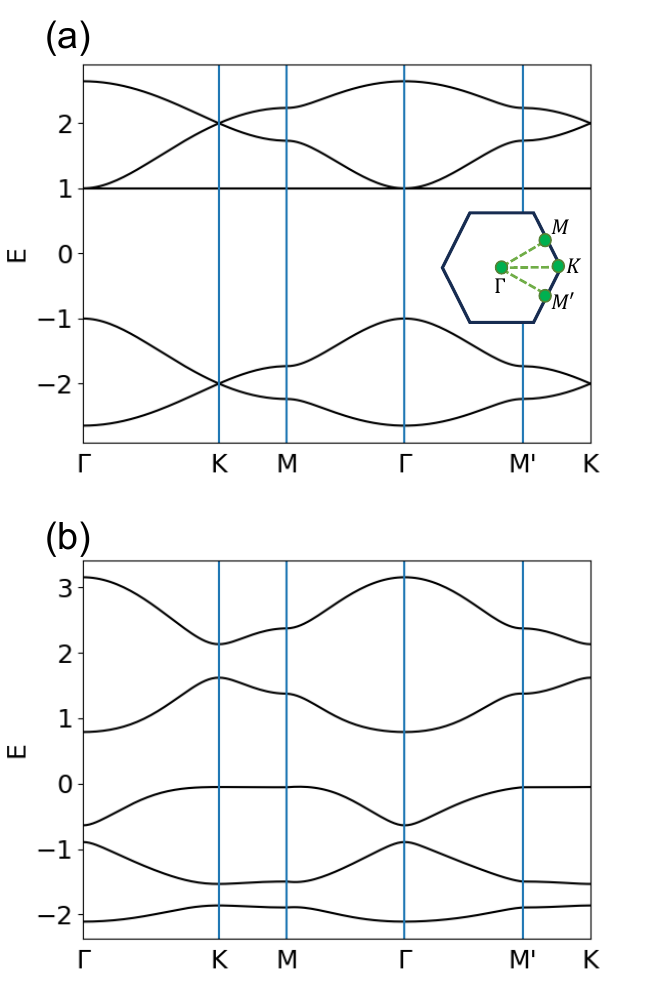}
  \caption{\label{fig:BK-band} Band dispersions of the Haldane model on the BK lattice in Eq.~\eqref{eq:h1}
  at (a) $M=1$ and $t_2=0$ and (b) $M=0.3$, $t_2=0.3$, and $\phi=0.3\pi$.
  The results are plotted along the high symmetric lines in the first Brillouin zone shown in the inset of (a).
  }
\end{figure}

\subsection{\label{subsec3-1}Bishamon-kikko lattice}
We first examine the band dispersions and the topological phase diagrams for the Haldane model on the BK lattice in Eq.~\eqref{eq:h1}.
Figure~\ref{fig:BK-band} shows the band dispersions along the high symmetric lines in the first Brillouin zone 
at (a) $M=1$ and $t_2=0$ and (b) $M=0.3$, $t_2=0.3$, and $\phi=0.3\pi$.

When the second-neighbor hopping is absent ($t_{2}=0$), 
a flat band appears at energy $E=M=1$, between the two sets of dispersive bands [Fig.~\ref{fig:BK-band}(a)]. 
This is associated with localized states at the yellow sites in Fig.~\ref{fig:schematic}(b), 
originating from quantum interference, as found in the Lieb lattice~\cite{PhysRevLett.62.1201}. 
For this parameter set, one of the upper dispersive bands has quadratic band touching with this flat band at the $\Gamma$ point. 
Each set of the dispersive bands shows linear band crossing at the $K$ point, forming the doubly-degenerate Dirac nodes like in the honeycomb lattice model.
Meanwhile, when the second-neighbor hopping is introduced,  
the flat band becomes dispersive, and the Dirac nodes at the $K$ point are gapped out, as shown in the result for $t_2=0.3$ and $\phi=0.3\pi$ in Fig.~\ref{fig:BK-band}(b).

\begin{figure}[thp]
 \includegraphics[width=\columnwidth]{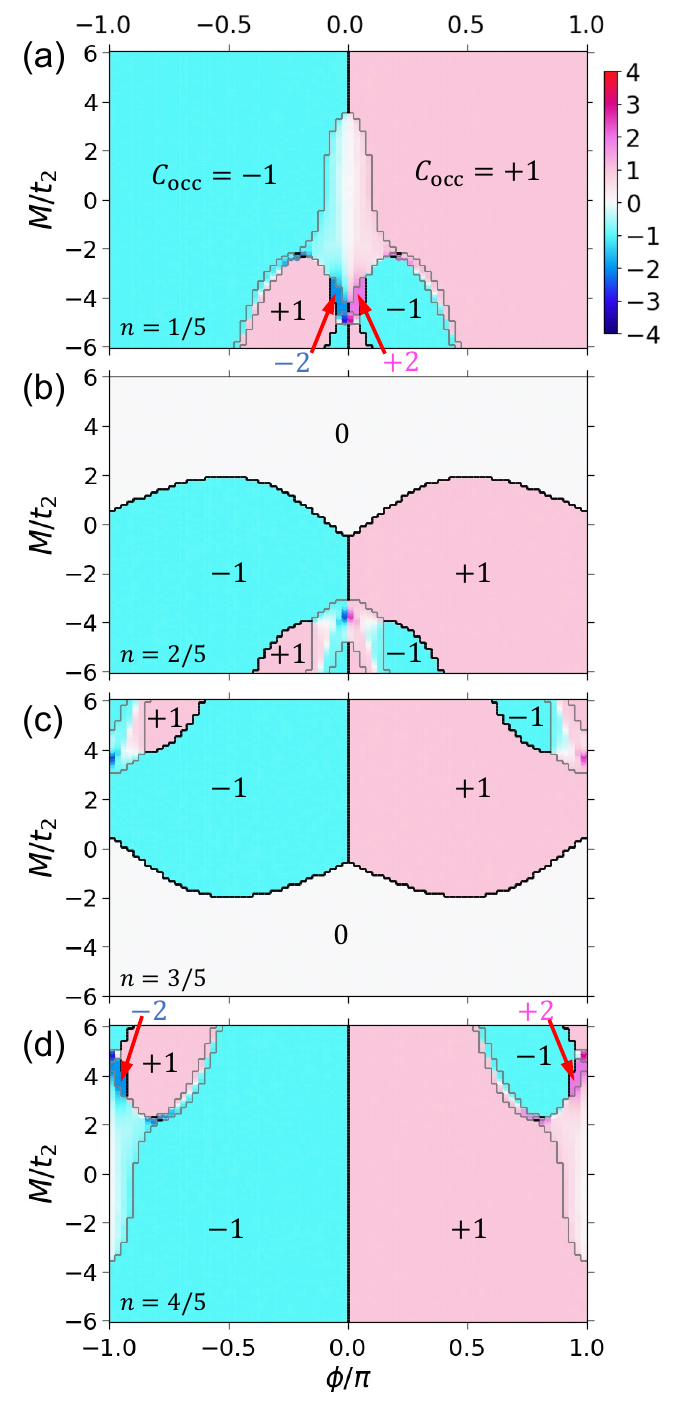}
  \caption{\label{fig:BK-pd} Topological phase diagrams of the Haldane model on the BK lattice in Eq.~\eqref{eq:h1} while changing $\phi$ and $M/t_2$. 
  The color contour denotes $C_{\rm occ}$ in Eq.~\eqref{eq:Chern}, the summation of the Berry curvature for the occupied states. 
  (a), (b), (c), and (d) correspond to 1/5, 2/5, 3/5, and 4/5 filling, respectively.
  The black lines represent the topological phase boundaries between the Chern insulators with different integers $C_{\rm occ}$,
  while the gray lines represent metal-insulator transitions where $C_{\rm occ}$ deviates from the integer values. 
 }
\end{figure}

Figure~\ref{fig:BK-pd} represents the topological phase diagrams at each commensurate filling for the Haldane model on the BK lattice.
The color denotes $C_{\rm occ}$ in Eq.~\eqref{eq:Chern}. 
The insulating regions with quantized integer values of $C_{\rm occ}$ are distinguished from the metallic regions with noninteger $C_{\rm occ}$ by the gray lines.
The black lines represent the phase boundaries between the insulators with different integers of $C_{\rm occ}$. 
There are several characteristic aspects in the phase diagrams, which are not seen in the original Haldane model on the honeycomb lattice at half filling~\cite{PhysRevLett.61.2015}. 
First of all, the phase diagrams are not symmetric with respect to the sign of $M/t_2$. 
This is because the BK lattice has different number of sublattice sites, unlike the honeycomb case [see Figs.~\ref{fig:schematic}(a) and \ref{fig:schematic}(b)]. 
Second, nevertheless, the phase diagrams are symmetric between $m$/5 and $1-m/5$ fillings ($m=1,2,3,4$) by changing the parameters as $\phi \rightarrow \phi+\pi$ and $M/t_{2} \rightarrow -M/t_{2}$. 
This is due to the particle-hole symmetry of the model with interchanging blue/yellow sublattices in Fig.~\ref{fig:schematic}(b). 
Indeed, the band structures are reversed in energy by the parameter changes (not shown). 
Third, most notably, in addition to the wide regions of topologically nontrivial insulators with $C_{\rm occ} = \pm 1$, which are also found in the honeycomb case, 
there appear another topologically nontrivial insulating phases with higher Chern number $C_{\rm occ}=\pm2$. 
Moreover, there exist metallic regions with metal-insulator transitions even at these commensurate fillings. 
These phases with high Chern numbers as well as the metal-insulator transitions are not found in the Haldane model on the honeycomb lattice, 
indicating that the Haldane model on the BK lattice has much richer topological phase diagrams compared with that on the honeycomb lattice.

We note that at $2/5$ ($3/5$) filling the system becomes a trivial insulator with $C_{\rm occ}=0$ for large (small) $M/t_2$, as shown in Fig.~\ref{fig:BK-pd}(b) [\ref{fig:BK-pd}(c)]. 
This is related to the fact that the model reduces to the honeycomb (kagome) lattice model in the limit of $M\rightarrow \infty \ (-\infty)$, 
as shown in Fig.~\ref{fig:schematic}(c) [\ref{fig:schematic}(d)]; the $2/5$ ($3/5$) filling for $M\rightarrow \infty$ corresponds to full filling of the honeycomb (kagome) model 
in Fig.~\ref{fig:schematic}(c) [\ref{fig:schematic}(d)], resulting in the trivial insulating states.

\subsection{\label{subsec3-2}Bishamon-kikko--honeycomb lattice}

\begin{figure}[thp]
  \includegraphics[width=\columnwidth]{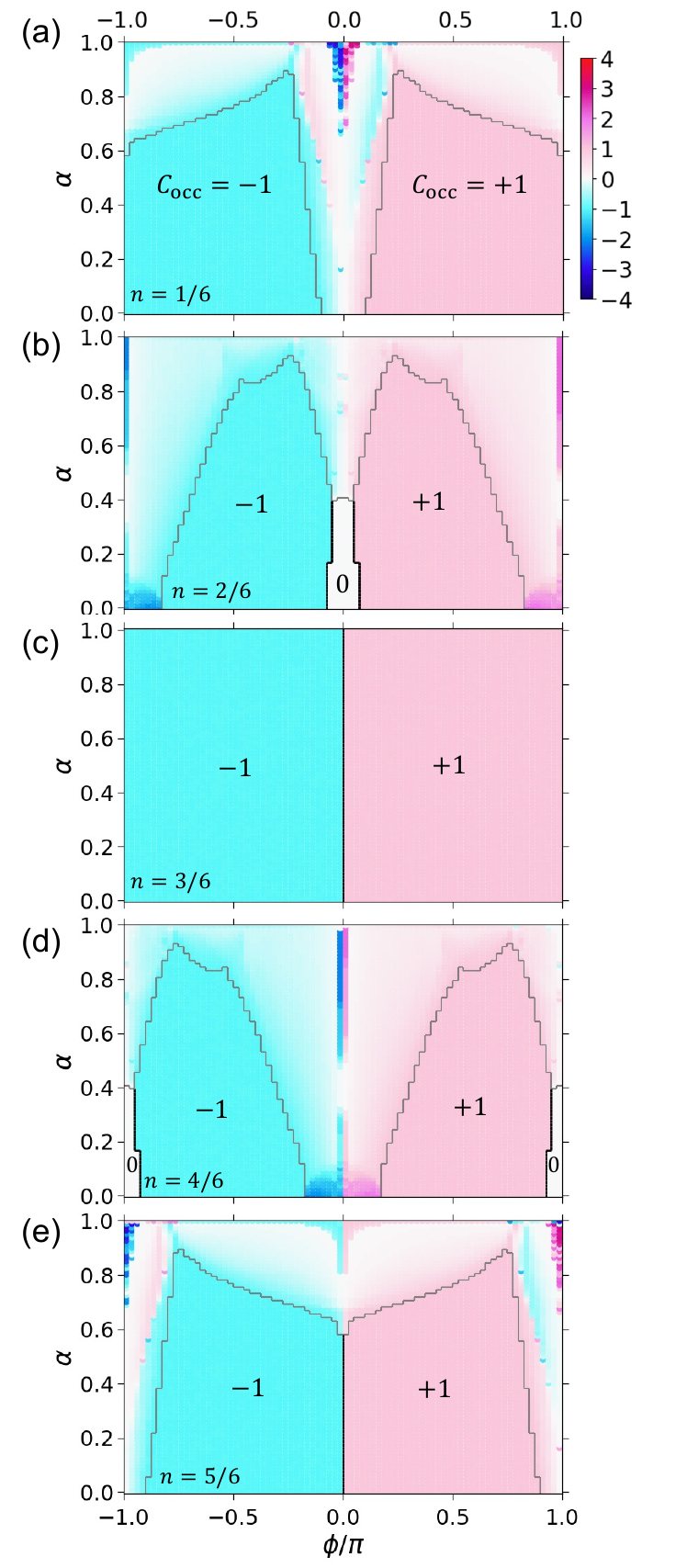}
  \caption{\label{fig:m0-pd} Topological phase diagrams of 
  the Haldane model on the BKH lattice in Eq.\eqref{eq:h2} 
  while changing $\phi$ and $\alpha$ at $M/t_{2}=0$. 
  $\alpha$ on the vertical axis denotes the parameter connecting 
  the BK lattice with isolated defects ($\alpha=0$) and the honeycomb lattice ($\alpha=1$).
  (a), (b), (c), (d), and (e) correspond to 1/6, 2/6, 3/6, 4/6, 5/6 filling, respectively. 
  The notations are common to those in Fig.~\ref{fig:BK-pd}.
}
\end{figure}

\begin{figure}[thp]
  \includegraphics[width=\columnwidth]{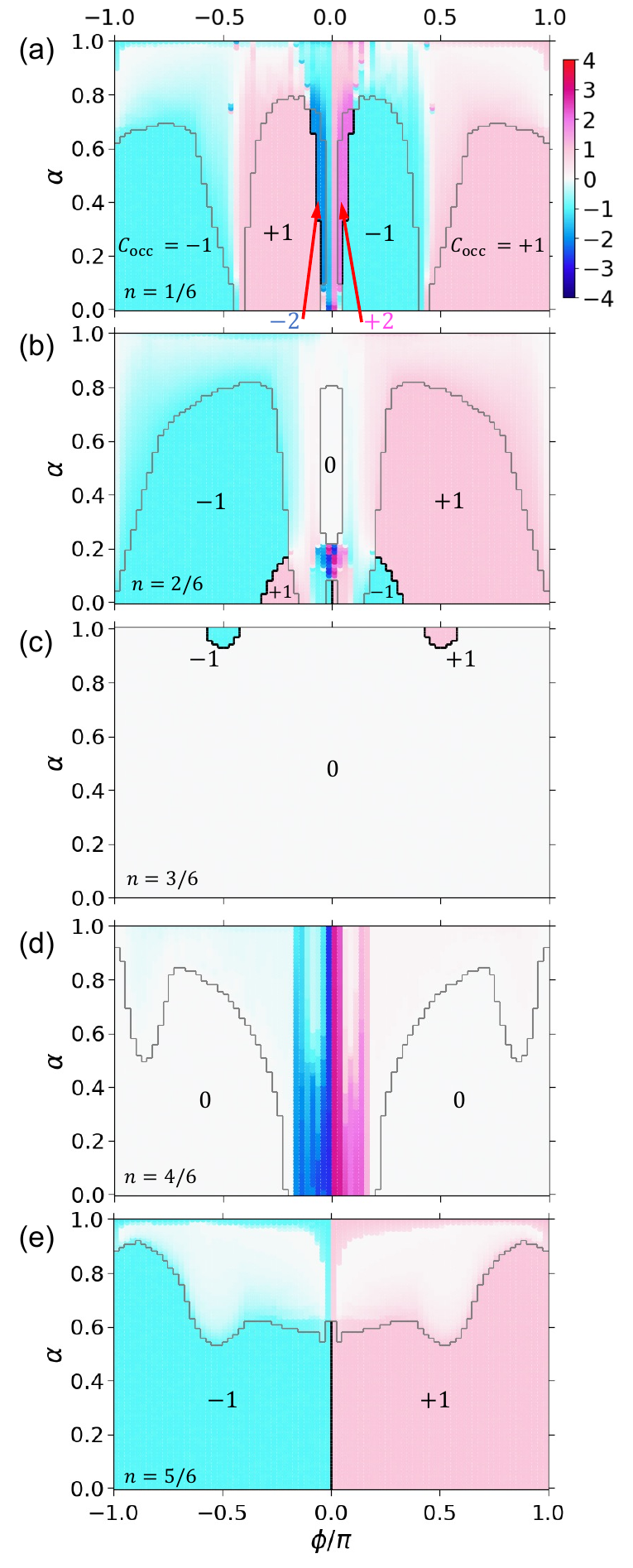}
  \caption{\label{fig:m-5-pd}  Topological phase diagrams of 
  the Haldane model on the BKH lattice in Eq.\eqref{eq:h2} 
  while changing $\phi$ and $\alpha$ at $M/t_{2}=-5$. 
  (a), (b), (c), (d), and (e) correspond to 1/6, 2/6, 3/6, 4/6, 5/6 filling, respectively. 
  The notations are common to those in Fig.~\ref{fig:m0-pd}.}
\end{figure}

Next we discuss the topological phase diagrams for the Haldane model on the BKH lattice in Eq.~\eqref{eq:h2}.
Among the model parameters $M$, $t_{2}$, $\phi$, and $\alpha$, 
we display the phase diagrams at each commensurate filling at particular values of $M/t_2$ with $t_2=0.3$ while varying $\phi$ and $\alpha$. 
The results for $M/t_{2}=0$ and $M/t_{2}=-5$ are shown in Figs.~\ref{fig:m0-pd} and \ref{fig:m-5-pd}, respectively.
The notations are common to those in Fig.~\ref{fig:BK-pd}.
The upper end at $\alpha=1$ in each panel corresponds to the honeycomb lattice,
while the lower end at $\alpha=0$ corresponds to the BK lattice with isolated defects.
Thus, the results for $\alpha=1$ at half filling ($3/6$ filling) shown in Figs.~\ref{fig:m0-pd}(c) and \ref{fig:m-5-pd}(c) reproduce those for 
the Haldane model on the honeycomb lattice at half filling with corresponding values of $M/t_2$~\cite{PhysRevLett.61.2015}. 

Let us first focus on the results for $M/t_{2}=0$ shown in Fig.~\ref{fig:m0-pd}.
As mentioned above, the lower end at $\alpha=0$ of each panel corresponds to the BK lattice with isolated defects; 
the values of $C_{\rm occ}$ at 1/6 and 5/6 fillings in Figs.~\ref{fig:m0-pd}(a) and \ref{fig:m0-pd}(e), respectively, 
are the same as those for $M/t_{2}=0$ at 1/5 and 4/5 fillings in Figs.~\ref{fig:BK-pd}(a) and \ref{fig:BK-pd}(d). 
In contrast, $C_{\rm occ}$ for $\alpha=0$ at 2/6, 3/6, and 4/6 fillings in Figs.~\ref{fig:m0-pd}(b), \ref{fig:m0-pd}(c), and \ref{fig:m0-pd}(d), respectively,
do not match with those for $M/t_{2}=0$ in Fig.~\ref{fig:BK-pd}. 
This is because the flat band at $E=-M$, originating from isolated sites on 
the BKH lattice, which locates at zero energy in this case, intervenes the dispersive bands.
We note that a topologically trivial insulator with $C_{\rm occ}=0$ appears in a narrow region sandwiched by the $C_{\rm occ}=\pm 1$ phases at $2/6$ filling in Fig.~\ref{fig:m0-pd}(b). 
While changing $\alpha$ from $0$ to $1$, we find that the Chern insulating phases with $C_{\rm occ}=\pm 1$ are shrunk gradually and 
taken over by metallic phases through metal-insulator transitions,
except for 3/6 filling in Fig.~\ref{fig:m0-pd}(c).
In this case, the insulating phases with higher integers $|C_{\rm occ}|$ do not appear, while high values are found in some metallic regions. 
We note that the model on the BKH lattice defined by Eq.~\eqref{eq:h2}
has a particle-hole symmetry between $m/6$ and $1-m/6$ fillings by changing  
$M \rightarrow -M$ and $\phi \rightarrow \phi+\pi$, similar to the BK lattice case; 
this is clearly seen in the results in Fig.~\ref{fig:m0-pd} with $M=0$.

Let us move on to the results for $M/t_{2}=-5$ shown in Fig.~\ref{fig:m-5-pd}.
As in the case with $M/t_{2}=0$ discussed above,
the values of $C_{\rm occ}$ for $\alpha=0$ at 1/6 and 5/6 fillings 
in Figs.~\ref{fig:m-5-pd}(a) and \ref{fig:m-5-pd}(e), respectively, match with those for $M/t_{2}=-5$ 
at 1/5 and 4/5 fillings in Figs.~\ref{fig:BK-pd}(a) and \ref{fig:BK-pd}(d), 
while not so at 2/6, 3/6, and 4/6 fillings, because of the intervening flat band for the isolated sites. 
Since $M\neq 0$, the particle-hole symmetry is not seen in these phase diagrams. 
Except for the $3/6$ and $4/6$-filling cases in Figs.~\ref{fig:m-5-pd}(c) and \ref{fig:m-5-pd}(d), respectively, 
the Chern insulating phases with $C_{\rm occ}=\pm 1$ are shrunk and taken over by metallic regions as in the case with $M/t_{2}=0$, 
but the phases with high Chern number $C_{\rm occ}=\pm 2$ appear in the 1/6-filling case, as shown in Fig.~\ref{fig:m-5-pd}(a).
We note that high Chern numbers $|C_{\rm occ}|$ are also found in the metallic regions, in particular, at the $1/6$ and $4/6$-filling cases. 

\section{\label{sec4}Summary}

To summarize, we have studied the topological phase diagrams in the Haldane model on the BKH lattice.
Our findings reveal several characteristics that distinguish this model 
from the original Haldane model on the honeycomb lattice:
topologically nontrivial insulating phases with high Chern number $C_{\rm occ}=\pm 2$, metallic regions including high $|C_{\rm occ}|$ even at commensurate fillings, 
and metal-insulator transitions between them.
Our results provide a basis for exploring intriguing phenomena induced by electron correlations, such as novel topological phases, magnetic ordering, and superconductivity. 
These lattice structures can be realized in a variety of two-dimensional materials, such as van der Waals materials~\cite{Wu2024,A.F.May2019}, 
graphene nanostructures~\cite{Bai2010,Zhang2017}, vacancy-engineered graphene \cite{PhysRevB.105.155414,PhysRevB.105.014511}, and photonic crystals~\cite{Lu2014,PhysRevB.80.155103}. 
Among them, the van der Waals materials introduced in Sec.~\ref{sec1} 
may offer the BKH lattice, 
when the interlayer hopping can be regarded as an effective in-plane hopping. 
Moreover, quantum spin systems on these lattices would also be intriguing because of the flat bands for quasiparticle excitations~\cite{Fukui2024}. 
 
\section*{\label{acknowledge}Acknowledgments}
We thank M. G. Yamada for valuable comments.
This work was supported by Japan Society for the Promotion of Science (JSPS) KAKENHI Grants No.~JP23H01119 and No.~JP24K17009.

\nocite{*}

\bibliography{reference}

\end{document}